\documentclass[fleqn,usenatbib]{mnras}

\usepackage[T1]{fontenc}

\DeclareRobustCommand{\VAN}[3]{#2}
\let\VANthebibliography\thebibliography
\def\thebibliography{\DeclareRobustCommand{\VAN}[3]{##3}\VANthebibliography}

\usepackage{graphicx}	
\usepackage{amsmath}	
\usepackage{amssymb}	

\newcommand{\Msunpc}{\,M$_\odot$\,pc$^{-1}$} 
\newcommand{\kms}{\,km\,s$^{-1}$} 
\newcommand{\K}{\,K} 
\newcommand{\gccm}{\,g\,cm$^{-3}$} 
\newcommand{\Myr}{\,Myr} 
\newcommand{\pc}{\,pc} 
\newcommand{\eg}{e.\,g.} 

\usepackage{newtxtext,newtxmath}

\title[Merging Filaments]{Merging Filaments I: A race against collapse}

\author[E. Hoemann et al.]{
Elena Hoemann$^{1}$\thanks{E-mail: hoemann@usm.lmu.de}, 
Stefan Heigl$^{1}$
and Andreas Burkert$^{1,2}$
\\
$^{1}$Universitäts-Sternwarte, Ludwig-Maximilians-Universität München, Scheinerstr. 1, 81679 Munich, Germany \\
$^{2}$Max-Planck Institute for Extraterrestrial Physics, Giessenbacherstr. 1, 85748 Garching, Germany
}

\date{Accepted XXX. Received YYY; in original form ZZZ}

\pubyear{2021}

\begin{document}
\label{firstpage}
\pagerange{\pageref{firstpage}--\pageref{lastpage}}
\maketitle

\begin{abstract}
  The interstellar medium is characterised by an intricate filamentary network which exhibits complex structures. These show a variety of different shapes (\eg\ junctions, rings, etc.) deviating strongly from the usually assumed cylindrical shape. A possible formation mechanism are filament mergers which we analyse in this study. Indeed, the proximity of filaments in networks suggests mergers to be rather likely. As the merger has to be faster than the end dominated collapse of the filament along its major axis we expect three possible results: (a) The filaments collapse before a merger can happen, (b) the merged filamentary complex shows already signs of cores at the edges or (c) the filaments merge into a structure which is not end-dominated. We develop an analytic formula for the merging and core-formation timescale at the edge and validate our model via hydrodynamical simulations with the adaptive-mesh-refinement-code RAMSES. This allows us to predict the outcome of a filament merger, given different initial conditions which are the initial distance and the respective line-masses of each filament as well as their relative velocities. 
\end{abstract}




\section{Introduction}\label{sec:Introduction}

  The cold interstellar medium (ISM) is characterised by elongated cylindrical density enhancements, known as filaments. Filaments are observed over many different scales, from $>100$\pc\ length like in Nessie \citep{Goodman2014, Mattern2018}, over other infrared dark clouds (IRDC) \citep{Perault1996, Alves1998, Egan1998,Hennebelle2001, Peretto2009, Miettinen2010}, down to the current detection limits of order of (sub-)pc scales \citep{Molinari2010, Arzoumanian2011, Hacar2013, Schmiedeke2021}. However, a clear understanding of their formation, evolution and collapse still remains an open question, although it has been known that they are a key step in the star formation process.

  Initial studies already revealed a close connection between filamentary structure and star formation \citep{Schneider1979}. Moreover, due to the Herschel dust observations \citep{Andre2010,Andre2014,Arzoumanian2019} combined with ground based molecular line observations \citep{Hacar2011, Kainulainen2016, Yuan2020} a much broader picture could be manifested in the last years. For example, most prestellar cores are found in supercritical filaments \citep{Koenyves2015, Andre2010} despite the fact that cores could also form by fragmentation of subcritical structures \citep{Nagasawa1987, Fischera2012,  Heigl2016, Gritschneder2017, Chira2018}.

  In low density environments, like in Taurus \citep{Hacar2013}, as well as inside dense regions, \eg in Orion \citep{Hacar2018}, small fialmentary substructure was observed insdie the large filaments. Thus, \citet{Tafalla2015} proposed a formation mechanism called `fray and fragment'. 
  The large filamentary structure is created due to an encounter of two gas fronts. Inside the filament velocity coherent structures form, which they called `fibers' these were also shown to be forming in turbulent, self-gravitating simulations \citep{Moeckel2015, Clarke2017}. However, \citet{Smith2014} also see fibers perpendicular to a filament, thus they proposed that these are not a product of filament fragmentation but rather accreted from the surrounding gas. Either way, this results in a dense filamentary network in which stars are supposed to form. The close proximity suggests that interactions between the fibers/filaments could occur and impact the further evolution of the region.
  
  In addition, observations show much more complex structures than the normally considered cylindrical shaped filaments. For example, in the substructure of the Orion integral filament, \citet{Hacar2018} found networks of small scale filaments where a common pattern appears to be the tuning-fork shape. These filaments reveal a split up of the single string into two components. Furthermore, even more complex configurations can be found, as \eg\ ring-like objects (closed filaments). The origin of these geometrical shapes could be related to the dynamics of the interstellar gas which is then of major importance for the physics of fragmentation and thus the early phase of stellar formation. 
  
  A scenario of tuning fork formation from an elliptical sheet in the context of the Orion A cloud has been proposed by \citet{Hartmann2007}. However, we explore a different formation mechanism: The merging of two cylindrical shaped filaments originally in hydrostatic equilibrium. It is an important mechanism to increase the density of a filament significantly and eventually trigger star formation. The fact, that in some observations multiple velocity components were detected inside a single filament \citep{Hacar2013, Yuan2020} could be a remnant of our scenario, as the filaments involved can retrain information about the merger in their intrinsic velocity field. Both \citet{Nakamura2014} (Serpens South) and \citet{Frau2015} (Pipe Nebula) have indeed detected signatures of colliding filaments and \citet{Fukui2021} stated that cloud-cloud collisions should lead to subsequent filament mergers. Nevertheless, theoretical work on the conditions of such a merger is currently very limited.

  However, the collapse timescale along the main axis of a filament limits the possibility of filament merging. Filaments collapse by the so called `edge effect'. Because the acceleration at the edges of the filaments is strongest \citep{Burkert2004}, matter is piled up and forms cores at the ends \citep[see observations by  \eg][]{Yuan2020,Cheng2021}. These then move into the centre while accreting the material on the way inwards and finally, fusing into one core and destroying the filament. Considering the case of filament mergers, this results in two competing timescales: the collapsing timescale of a filament \citep{Toala2011, Pon_2012, Clarke2015} versus the filament-filament merging timescale. A comparison of these two timescales enables us to determine under which conditions a filament can merge before the collapse destroys it. Since not all observed tuning-forks show signs of an end dominated collapse it is also interesting to investigate on which timescale cores could form at the ends, to get a more accurate idea of initial conditions required to form such structures. Thus, we explore analytic models for merging and edge effect formation timescales to explore the parameter space for merging filaments, initially in hydrostatic equilibrium.

  This paper is organised as follows: After an overview over the basic principles which were applied in our analysis (Section \ref{sec:Basic Principles}), the simulation framework is introduced (Section \ref{sec:Numerical setup}). The main part of the work focuses on the calculation of the merging (Section \ref{subsec:MergingTimescale}) and edge effect formation timescales (Section \ref{subsec:EdgeEffect}), each validated by simulations, respectively. This leads to the initial conditions under which mergers can take place, discussed in Section \ref{subsec:ConditionsForAMerger}. Finally, the results are discussed in Section \ref{sec:discussion} and conclusions are drawn in Section \ref{sec:conclusion}.


\section{Basic Principles} \label{sec:Basic Principles}

  We consider filaments to be isothermal gas cylinders in hydrostatic equilibrium, a model which has been investigated by  \citet{Stodolkiewicz1963} and \citet{Ostriker1964}. Both found that the radial profile follows:
  \begin{align} \label{eq:OstrikerProfile}
    \rho (r) = \rho_c \cdot \left[ 1 + \left( \frac{r}{H} \right)^2 \right]^{-2}
  \end{align}
  $H$ is the scale height, given by:
  \begin{align}
    H^2=\frac{2c_s^2}{\pi G\rho_c}
  \end{align}
  with $\rho_c$ being the central density of the filament, $c_s$ the sound speed which is 0.19\kms\ for $T=10$\K\ and a mean molecular weight of 2.36 as in \citet{Fischera2012}. We constrain the radius by the external pressure of the ISM, as the observed filaments do not extend to infinity \citep{Fiege2000}. 
  This external pressure $P_{\text{ext}}$ then sets the boundary density $\rho(R)=\rho_b$ of the filament such that it is in pressure equilibrium with the surrounding $P_{\text{ext}}=P_b$ which gives the filament a finite radius.

  A characterising quantity of the filament is the so called line-mass, the mass per length:
  \begin{align}
	\mu = \frac{M}{L}
  \end{align}
  The solution with the largest stable line-mass is then given by integrating the Ostriker-profile until infinity. For even larger line-masses the filament will collapse due to gravity as no hydrostatic solutions exist. This is called the critical line-mass $\mu_{\text{crit}}$. Considering the values mentioned above leads to:
  \begin{align}
	\mu_{\text{crit}} = \left(\frac{M}{L}\right)_{\text{crit}} = \frac{2c_s^2}{G} \approx 16.4 \text{\Msunpc}
  \end{align}
  To define the criticality of a filament, the parameter $f$ was introduced \citep{Fischera2012} as the ratio of the filament line-mass to the critical line-mass:
  \begin{align}
	f = \frac{\mu}{\mu_{\text{crit}}}
  \end{align}
  which allows to connect the central and boundary density:
  \begin{align}\label{eq:rho_b}
	\rho_b = \rho_c \cdot (1-f)^2
  \end{align} 
  Analogous the radius of the filament is given by:
  \begin{align}
	R = H \left( \frac{f}{1-f} \right)^{1/2}
  \end{align}

  Considering now that filaments in the ISM have a finite length, they will collapse along their major axis under self-gravity, when there are no other external influences \citep{Keto2014}. The acceleration along such a filament of length $l$ was investigated by \citet{Burkert2004} and is given by:
  \begin{align} \label{eq:accelerationCollapseFilament}
	a = -2 \pi G \rho \left[ 2z - \sqrt{\left( \frac{l}{2} + z \right)^2+R^2} + \sqrt{\left( \frac{l}{2} - z \right)^2+R^2} \right]
  \end{align}
  where $z$ is the coordinate along the filaments main axis, with $z=0$ at the symmetry point. The acceleration shows a strong increase at the filament's edges. This leads to end-dominated collapse 
  and to core formation at the ends of the filament, the so called `edge effect' which, finally, destroys the filament. The lifetime of a filament is then given by the time it takes to collapse into a single core. As already mentioned, there exist several investigations of this collapse timescale \citep{Toala2011, Pon_2012}, the latest by \citet{Clarke2015}, which is given by:
  \begin{align}
    t_{\text{col}} = \frac{0.49 + 0.26 \, A}{\sqrt{G \rho}}
    \label{eq: Clarke}
  \end{align}
  Here, $A=l/(2R)$ is the aspect ratio and $\rho$ its average density $\mu /(\pi R^2)$.

\section{Simulation Setup} \label{sec:Numerical setup}

  We also compare our analytic approach to simulations which were executed using the adaptive-mesh-refinement code RAMSES, developed by \citet{Teyssier_2002}. It solves the Euler Equations in their conservative form by using a second-order Gudonov solver. We applied the MUSCL \citep[Monotonic Upstream-Centered Scheme for Conservation Laws,][]{Leer1979}, the HLLC-Solver \citep[Harten-Lax-van Leer-Contact,][]{Toro1994} and the MC slope limiter \citep[monotonized central-difference,][]{Leer1979}. The grid applied in our simulations varied from level 7, $(128)^3$ cells ($4.7\times 10^{-3}$\pc\,-\,$11.7 \times 10^{-3}$\pc\ depending on the box size), to level 9, $(512)^3$ cells ($1.2\times 10^{-3}$\pc\,-\,$2.9 \times 10^{-3}$\pc), such that the external medium is only resolved with low resolution while the filament itself is resolved with the highest one. As an example, a simulation as performed for Section \ref{subsec:MergingTimescale} has a resolution of 44 cells along the filaments diameter at its highest contraction and above 100 for its initial configuration. In order to resolve the hydrodynamical processes the filaments were resolved with a minimum of $\sim20$ cells.

  Two different types of simulations were carried out to verify the merging and edge effect formation timescales. The initial physical conditions used in each case are described in the corresponding section. Nevertheless, all simulations were carried out in the same environment, which means with the same boundary density of $\rho_b = 1.92 \times 10^{-20}$\gccm\ which equals $4.9 \times 10^3$ particles per cm$^3$, a rather high value to recreate a surrounding as in Orion. Given $f$ this also sets the central density via Equation \ref{eq:rho_b}. This resembles the situation where all filaments are embedded in the same surrounding medium, constrained by the same outside pressure. However, tests showed that variations in the external pressure does not have a major influence on the result of the simulation. The external density was set to $\rho_{\text{ext}} = 3.92 \times 10^{-23}$\gccm, in pressure equilibrium to the filament, in order to minimise accretion effects.


\section{Merging timescale} \label{subsec:MergingTimescale}

  In this section the merging process of two filaments is described in detail. This includes the calculation of the merging timescale and the determination of the trajectory.
  \begin{figure}
	\centering
	\includegraphics[width=1.0\columnwidth]{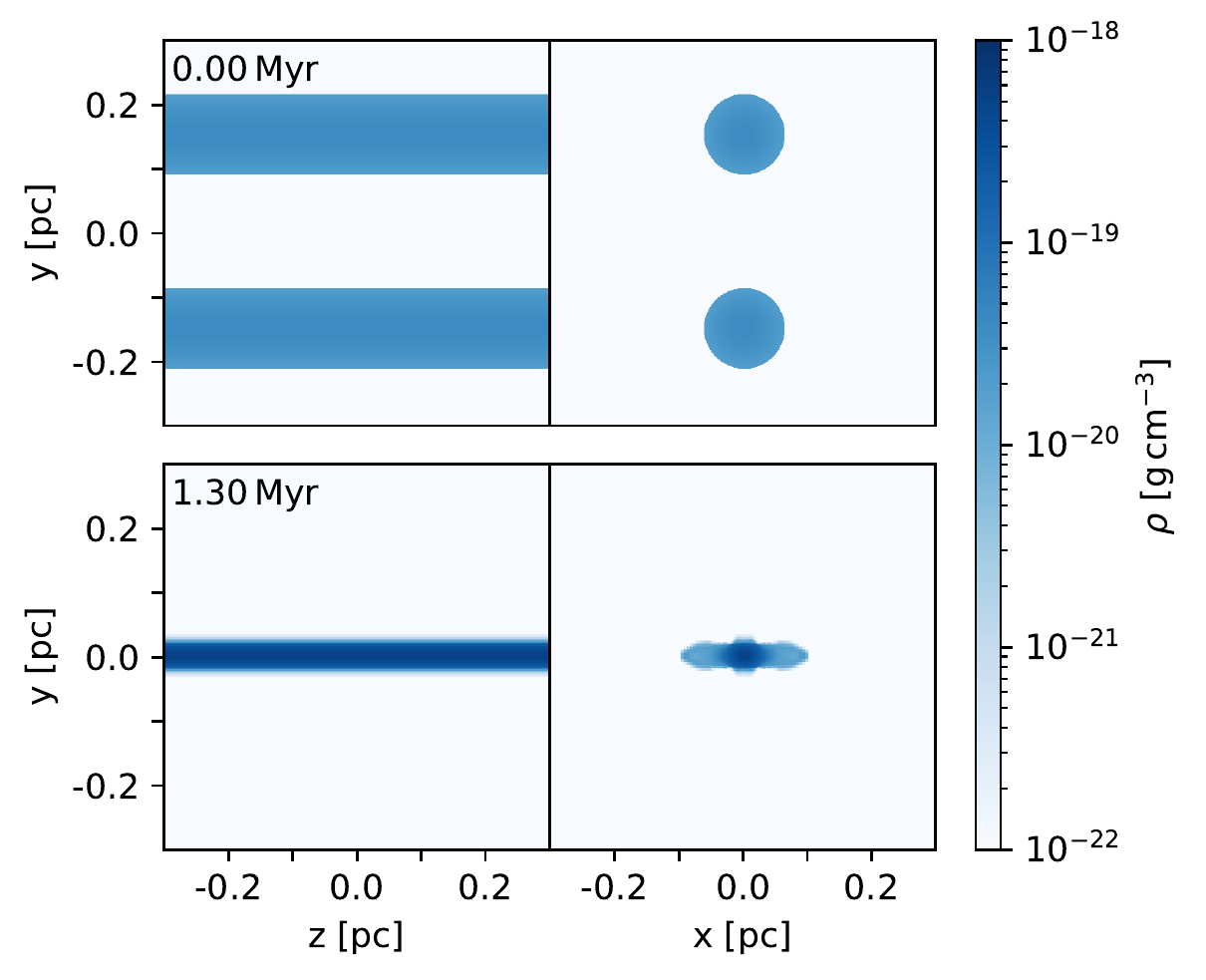}
	\caption{Density slices of a `standard' simulation done in this work $f_1=f_2=0.3$ of merging filaments. In the first row the initial condition ($t=0.00$\Myr) and in the second row the time of final configuration ($t=1.30$\Myr) is displayed. The left hand side shows the the filament elongated in z direction and initially separated in y direction (z/y). On the right hand side the slice is given in x/y-direction to see the whole dimensionality of the simulation.}
	\label{fig:Simulation_Snapshots}
  \end{figure}
  \begin{figure} 
	\centering 
	\includegraphics[width=1.0\columnwidth]{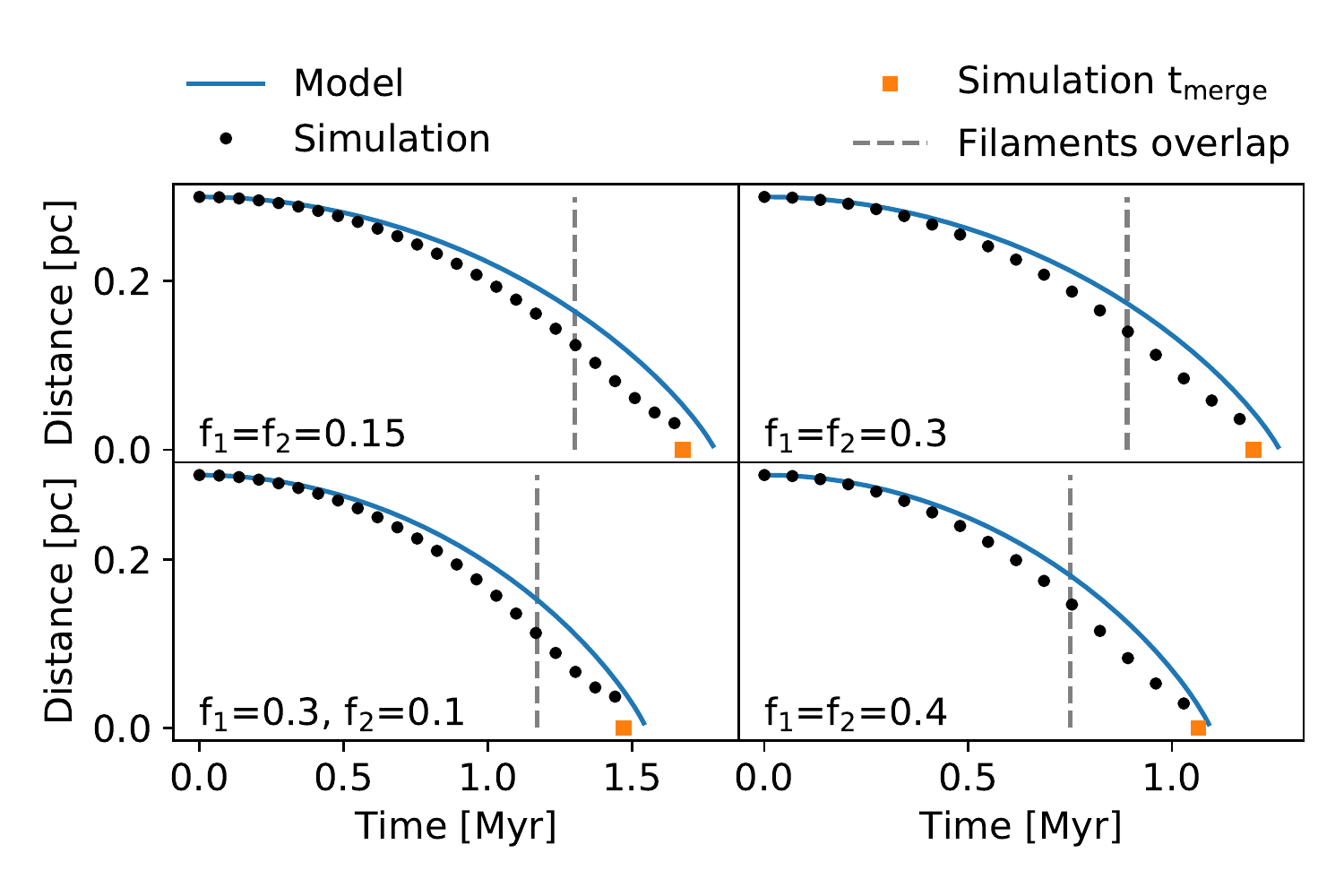}
	\caption{Validation of the trajectory of merging filaments with four different constellations of line-masses ($f_{1}$, $f_{2}$ are given in the bottom left corner). The blue curve shows the trajectory given by Equation \ref{eq:trajectory_merging}, the black dots indicate the centre of mass distance of the filaments. The grey dashed line represents the moment when the filaments start to overlap. The orange square gives the simulation's merging time. Given the simplicity of the analytical approach we consider this a good agreement between the values from simulation (dots) and the model.} 
	\label{fig:validation_merging_timescale}
  \end{figure}
  In order to derive the merging timescale, two infinite filaments with according line-masses of $\mu_1$ and $\mu_2$ are considered to be in a parallel configuration of initial distance $d_0$, as depicted in Figure \ref{fig:Simulation_Snapshots} in the first snapshot.

  The gravitational acceleration on both filaments respectively is given by
  \begin{align}\label{eq:gravAttaction}
    a_G = - \left( \mu_1+\mu_2 \right) \frac{2G}{d}
  \end{align} 
  which results in the following relative velocity, with $v_0$ being the initial relative velocity of the filaments:
  \begin{align} \label{eq:velocity}
    v = - \sqrt{4G\left( \mu_1 + \mu_2 \right) \log \left( \frac{d_0}{d} \right) + v_0^2}
  \end{align}
  Integration from the initial separation $d_0$ to $d$ results in the time needed for the filaments to reach the distance $d$ (derivation in Appendix \ref{ap: Calculation of the merging timescale}):
  \begin{align} \label{eq: t merger t(r)}
    t =& \sqrt{\frac{\pi}{G(\mu_1+\mu_2)}} \cdot \frac{d_0}{2} \cdot \exp \left( \frac{v_0^2}{4G(\mu_1+\mu_2)} \right) \\
    & \cdot \left[ \mathrm{erf} \left( \sqrt{\log \left( \frac{d_0}{d} \right) + \frac{v_0^2}{4G(\mu_1+\mu_2)} } \right) -\mathrm{erf} \left( \frac{v_0}{\sqrt{4G(\mu_1+\mu_2)}} \right) \right] \nonumber
  \end{align}
  with $\mathrm{erf}(x)$ being the Gauss error function. In the limit of $d$ approaching zero, which means that the filaments overlap completely, the merging time can be approximated:
  \begin{align} \label{eq: merging_time}
    t_{\text{merge}} =& \lim\limits_{d\rightarrow 0} t 
    \nonumber \\
    =& \sqrt{\frac{\pi}{G(\mu_1+\mu_2)}} \cdot \frac{d_0}{2} \cdot \exp \left(\frac{v_0^2}{4G(\mu_1+\mu_2)}\right) 
    \nonumber \\
    &\cdot \left[ 1 - \mathrm{erf} \left( \frac{v_0}{\sqrt{4G(\mu_1+\mu_2)}} \right) \right]
  \end{align}
  If the filaments are initially at rest, $v_0=0$, the merging time reduces to:
  \begin{align}
    t_{\text{merge}} = \sqrt{\frac{\pi}{G(\mu_1+\mu_2)}} \cdot \frac{d_0}{2}
  \end{align}
  For $v_0=0$ the trajectory, distance of the filaments depending on time, is then given by solving Equation \ref{eq: t merger t(r)} for $d$:
  \begin{align} \label{eq:trajectory_merging}
    d(t) = d_0 \cdot \exp \left[ - \mathrm{erfinv} \left( 2 \sqrt{\frac{G(\mu_1+\mu_2)}{\pi}} \frac{t}{d_0} \right)^2 \right]
  \end{align}
  with $\mathrm{erfinv}(x)$ being the inverse error function. 

  In order to validate these results, we execute simulations of filament mergers with different line-masses and initial velocities. The simulation box has a scale of (0.6\pc)$^3$. The initial condition is given in Figure \ref{fig:Simulation_Snapshots} in the first row: the left hand side shows a density slice in z-y direction and the right hand side a density slice in x-y direction, in order to display the full dimensionality. We present two points in time, the initial condition $t=0.0$\Myr\ and the merged filaments, in this example $t=1.3$\Myr. The filaments have their major axis along the z-axis which has periodic boundary conditions in order to create an infinitely long filament. They are initially separated by $d_0=0.3$\pc\ in y-direction. The boundary in y and x direction were chosen to be open. During the simulation the two filaments fall into each others potential and merge. To exclude that accretion has major effects on the simulation we varied the external density ($\rho_{\text{ext}} = 3.92\times 10^{-24}$\gccm) but no influences on the merger were observed. 

  A comparison of theoretical and simulated distance evolution of the two filaments for different merging constellations is given in Figure \ref{fig:validation_merging_timescale}. The theoretical distance prediction is depicted in blue for different line-masses. In contrast, the black dots represent the distances obtained from simulations by determining the distance of the centre of mass of the two filaments. The gray dashed line indicates the time the filaments start to overlap. The simulations (black dots) show a very good agreement to the theoretically predicted values (blue line). The orange square represents the merging time determined from simulations, which should be compared to the zero point of the trajectories. The merging point of the simulation was considered to be the time where the merged filament recovers the radius of the initial filaments, in this case the two initial filaments totally overlap and share the same centre of mass ($d=0$). For the merger with different line masses and thus different radius, we used the mean of the filaments radius instead. In all cases we get a good agreement between the model and the simulated values.

  In order to confirm that the model correctly reproduces the merging time's velocity and line-mass dependence, we also performed a parameter study. 
  \begin{figure}
	\centering
	\includegraphics[width=1.0\columnwidth]{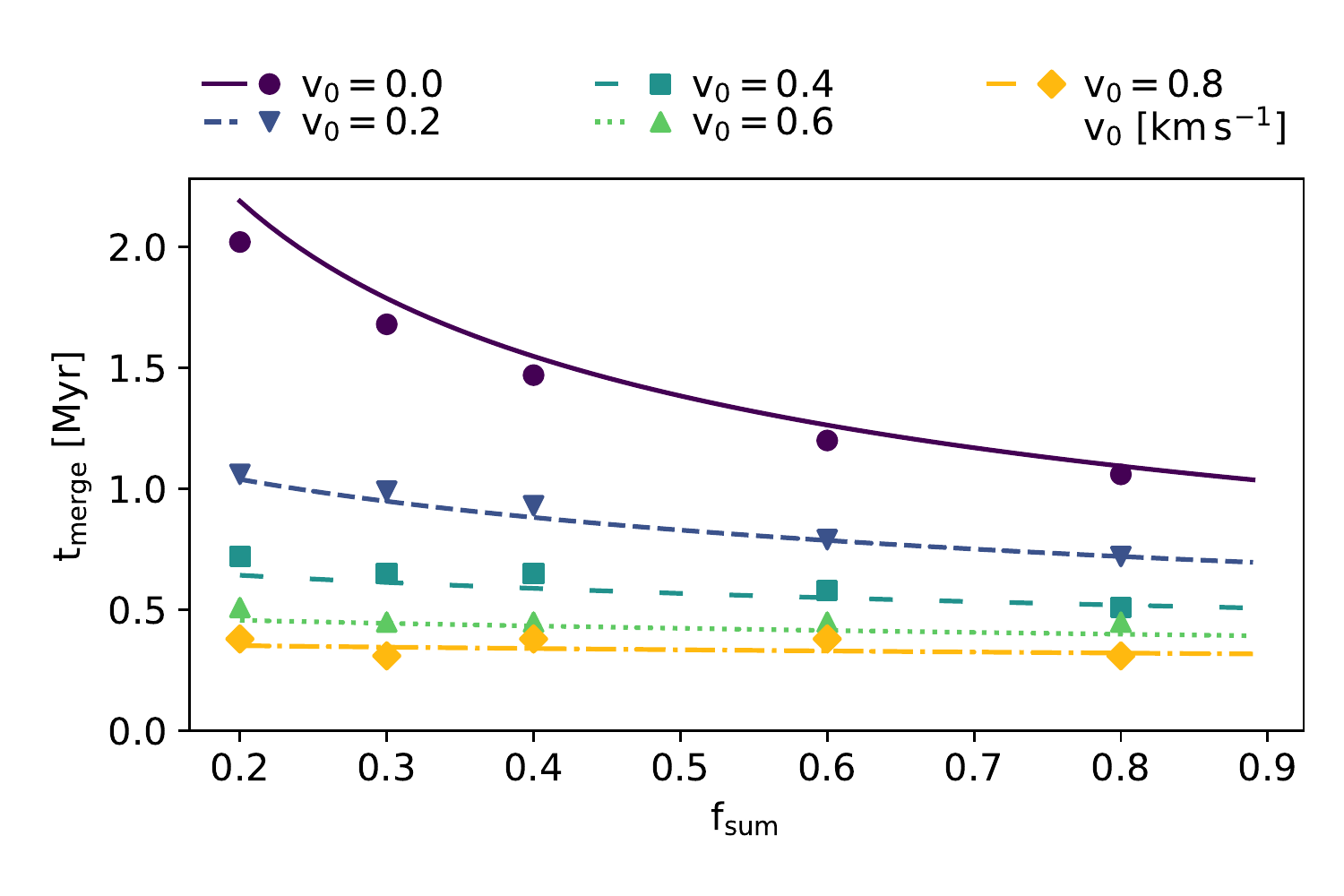}
	\caption{Validation of the merging timescale depending on the sum $f_{\text{sum}}=f_1+f_2$. The different coloured lines show Equation \ref{eq: merging_time} for different initial velocities but fixed $d_0=$ 0.3\pc. The values from the simulation are displayed as symbols in the corresponding colour.}
	\label{fig:MergingTimescale}
  \end{figure}
  The results are shown in Figure \ref{fig:MergingTimescale}. The different coloured lines are the solutions to Equation \ref{eq: merging_time} for different initial velocities. The symbols show the corresponding merging time of the simulation, which is in good agreement with the predictions. Both, the dependence on the summed line-mass, as well as on the initial relative velocity of the two filaments, $v_0$, are reproduced very well in our simulations.
  %


\section{Edge effect formation timescale}\label{subsec:EdgeEffect}

  If we consider now a more realistic setup with finite filaments a second effect comes into play: the gravitational collapse of the filaments along their major axis. This leads to the important question of which process is faster, the merger or the edge effect. The collapse time was already investigated by \citet{Clarke2015} which we  discussed in Section \ref{sec:Basic Principles}. However, as not all observed tuning-forks show signs of the edge effect we want to investigate on what timescale cores are formed at the edges which we will call the edge effect formation timescale $t_{\text{edge}}$. This can give us constraints on the initial conditions of the merged filaments.

\begin{figure}
	\centering
	\includegraphics[width=1.0\columnwidth]{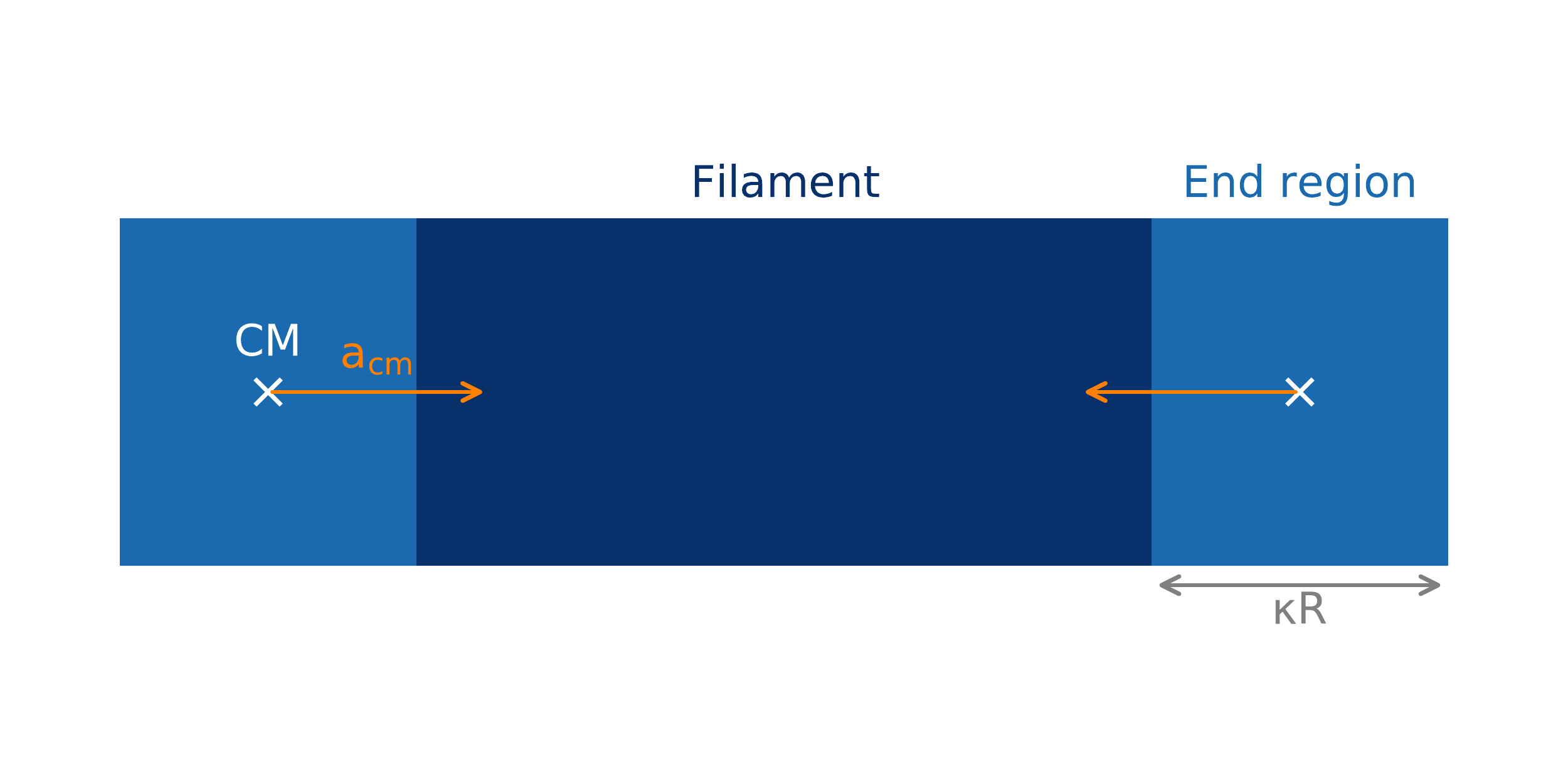}
	\caption{Schematic drawing of a face-on filament. With two end regions of size $\kappa R$ (light blue areas) which accumulate the mass. The centre of mass of the end regions is marked with white crosses. Here the acceleration $a_{\text{cm}}$ is indicated by the orange arrow.}
	\label{fig:ColapseTimescaleScheme}
  \end{figure}

  We consider the filament to have two end regions, similar to \citet{Yuan2020}. A simple scheme is given in Figure \ref{fig:ColapseTimescaleScheme} where these end regions are marked with light blue color. They are considered to be the zone where the end clump will evolve.
  The total acceleration of this region is approximated to be the acceleration of the end regions centre of mass, indicated by the white cross with attached orange arrow depicting the acceleration. For the right edge located at $z=\frac{l}{2}-\frac{\kappa R}{2}$ it is given by Equation \ref{eq:accelerationCollapseFilament}, for $l \gg R$: 
  \begin{align}
    a_{\text{CM}}\left( \frac{l}{2} - \frac{\kappa R}{2} \right) \approx - \pi G \rho R \left( \sqrt{\kappa^2+4} -\kappa \right)
  \end{align}
  The constant value of their extension $\kappa R$ is afterwards fitted to the simulations. In case of a spherical end region one would expect its size to be $2R$, thus $\kappa=2$. The acceleration can be considered as constant in time as long as $l \gg R$ holds, because the most time dependent parameter (length $l$) cancels out. We assume the end region to accumulate all the mass which it encounters during the collapse:
  \begin{align}
    M (t) = M_0 + \frac{1}{2} a_{\text{CM}} t^2 \cdot \mu_0
  \end{align}
  This leads to the time dependent line-mass of the end region:
  \begin{align}
    \mu (t) = \frac{\kappa R+\frac{1}{2}a_{\text{CM}}t^2}{\kappa R} \cdot \mu_0
  \end{align}
  The timescale of interest is the time needed for the end region to get supercritical:
  \begin{align}
    f(t) = \frac{\mu (t)}{\mu _{\text{crit}}} \stackrel{!}{=} 1
  \end{align}
  with $\mu_{\text{crit}}$ the critical line-mass. From here we are able to calculate the timescale needed for the filament to accumulate supercritical end regions:
  \begin{align}\label{eq:CollapseTime1.0}
    t_{\text{edge},1.0} &= \sqrt{\left( \frac{1}{f} -1 \right) \frac{2\kappa R}{|a_{\text{CM}}|}} \\
    &= \sqrt{ \frac{(f-1)^2}{f} \cdot \frac{2}{\pi G \rho_b \left( \sqrt{1+(2/\kappa)^2} - 1 \right) }} 
  \end{align}
  In the context of filament mergers, another interesting timescale is the time necessary for the edges to be above a criticality of $f=0.5$. If there is a symmetrical merger the end regions of both filaments will overlap and thus if both end regions have a criticality of 0.5 they immediately become supercritical. The timescale in this case is given by:
  \begin{align}
    t_{\text{edge},0.5} = \sqrt{\left( \frac{1}{f} -2 \right) \cdot \frac{1-f}{\pi G \rho_b \left( \sqrt{1+(2/\kappa)^2} - 1 \right) }}
  \end{align}
  \begin{figure}
	\centering
	\includegraphics[width=1.0\columnwidth]{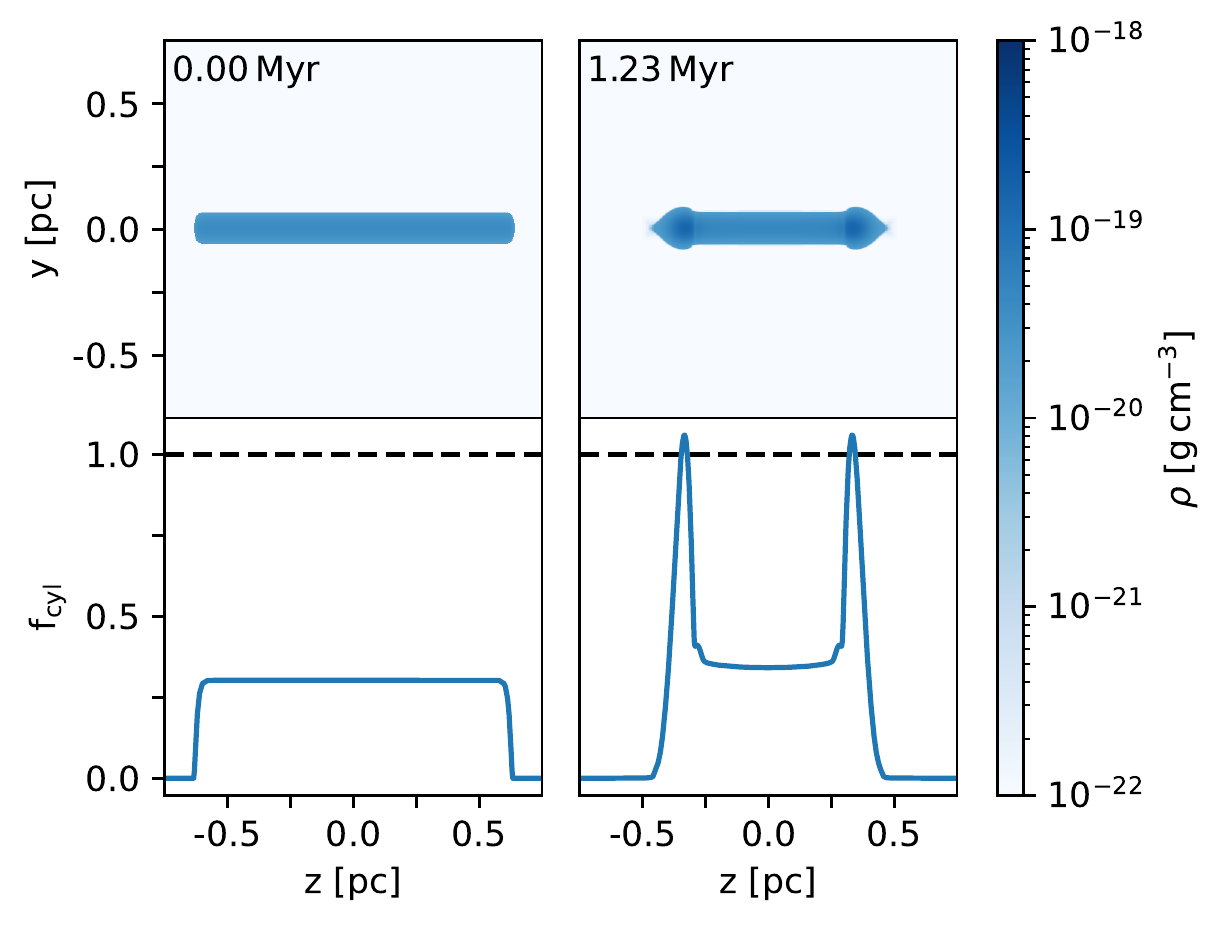}
	\caption{Top: Density slices of a `standard' simulation with $f=0.3$ to determine the edge formation timescale are presented in z-y direction. Bottom: The criticality of the filament along the z-axis is displayed. Left panel: The initial condition ($t=0.00$\Myr) of the simulation. Right panel: A snapshot after 1.23\Myr\ where the cores reach a criticality above 1 and thus start to collapse.}
	\label{fig:Simulation_Snapshots_collapse}
  \end{figure}

  This simple approach does not consider ram pressure contributions from the gas inside the filament. \citet{Clarke2015} showed that this effect was important when determining the overall collapse timescale of the filament. Following their semi-analytical model we found ram pressure to be only significant in the late stages of filament collapse and has only minor influences on the edge formation timescale. However for low mass filaments ($f\sim0.1$), having long edge formation timescales, this could get essential. Thus, we excluded all filaments with $f<0.2$ from our analysis.

  In addition, we also validated our results with simulations. We set up a filament in a box with a size of 1.5\pc\ and open boundary conditions. Thus, the filaments had a finite length of approximately 1.3\pc. We used an exponential cutoff for the filaments edge following $[1+\exp{(-100\cdot(0.4-z^2))}]^{-1}$. However, tests of different edges with sharp and soft cutoffs showed no influence on the simulated collapse time. The initial and final conditions are presented in Figure \ref{fig:Simulation_Snapshots_collapse}. The upper panel shows the density slices in z/y-direction at $t=0$ and $t=t_{\text{edge},1.0}$ and the bottom panel depicts the corresponding line-mass dependency along the z-axis. In the beginning, the filament has a uniform criticality of $f=0.3$ and we determine $t_{\text{edge},1.0}$ when the line-mass in the cores at the edge exceeds the limit of $f=1.0$ for the first time, as shown in the dashed black line in Figure \ref{fig:Simulation_Snapshots_collapse}. From the figure, the large density and line-mass enhancements at the edges are obvious. 
  The large volume of the box was chosen such that the condition $l\gg R$ holds. Therefore, the acceleration discussed in Section \ref{sec:Basic Principles} can be considered constant. We set a fixed boundary density of $\rho_b = 1.92\times 10^{-20}$\gccm\ and adjusted the central density (Equation \ref{eq:rho_b}) as before. The external density was again set to $\rho_{\text{ext}} = 3.92\times 10^{-23}$\gccm.

\begin{figure}
	\centering
	\includegraphics[width=1.0\columnwidth]{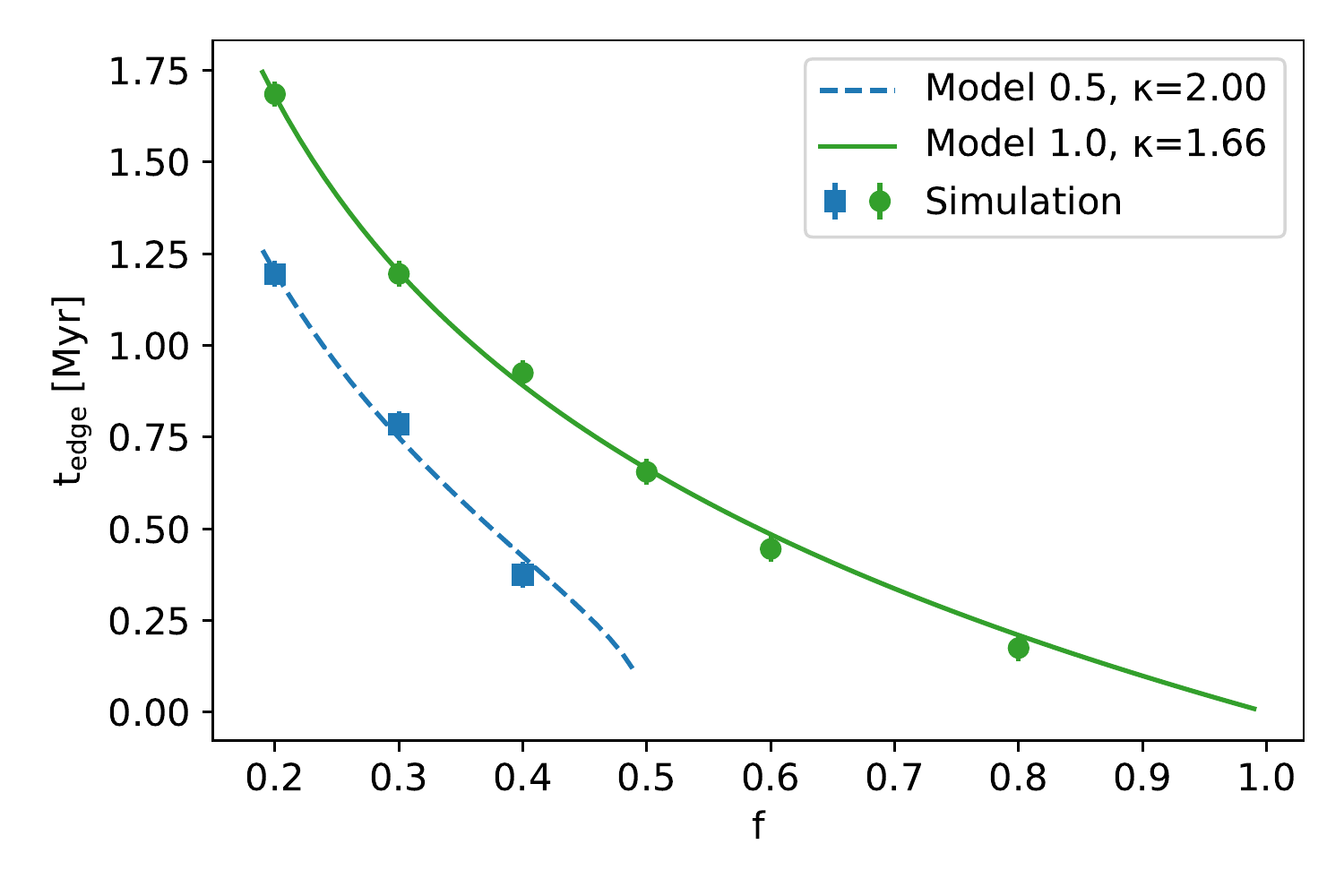}
	\caption{The edge effect formation timescale of the model as function of the initial $f$ (lines) in comparison to the simulation (symbols) is given. Green represents the model with maximum $f_{\text{max}}=1.0$ and blue $f_{\text{max}}=0.5$. The size of the corresponding end region $\kappa$ is fitted to the simulation data, the fit values are given in the legend. Both models show excellent agreement to the simulation.}
	\label{fig:ColapseTime_validation}
  \end{figure}

  Figure \ref{fig:ColapseTime_validation} shows the theoretical models (lines) of the core-formation timescale depending on $f$ in comparison to the values determined by our simulations (symbols). 
  Model $0.5$ (blue dashed line and squares) gives the timescale needed for a filament to form cores at the edges with core criticality $f=0.5$. Model $1.0$ (green solid line and dots) shows the timescale to grow cores with $f=1.0$. The size of the end region $\kappa R$ was fitted to the simulation data and the parameters are provided in the legend of the plot. As expected the deviation from $\kappa=2$ is small which considers a spherical end region. By inserting all constants ($\kappa$,$G$,$\pi$) and using Equation \ref{eq:rho_b} we have found a simple empirical relation that fits these cases very well: 
  \begin{equation} \label{eq: emp. model}
    t_{\text{edge},1.0} = \sqrt{\frac{1.69\times 10^{-20}\text{\gccm}}{f\rho_c}} \text{\Myr}
  \end{equation}
  Thus, the edge effect formation time only depends on the central density and the criticality of the filament.
  As Equation \ref{eq: emp. model} agrees well with the simulations we will use it  for the early collapse in order to constrain our initial conditions for a merger.
  
  \begin{figure}
	\centering
	\includegraphics[width=1.0\columnwidth]{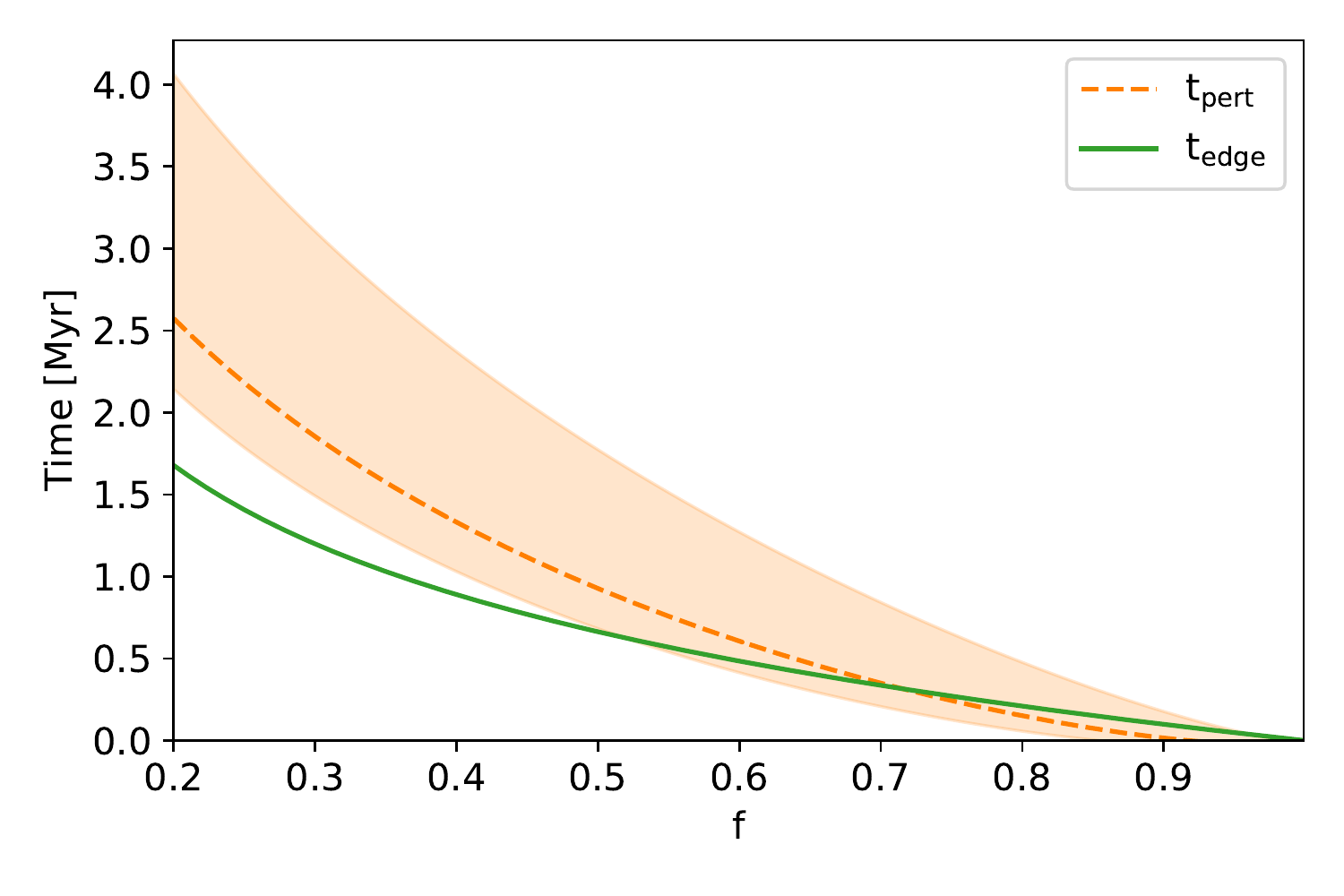}
	\caption{The core formation timescale (orange dashed line) for a perturbation strength of $0.09$ (light orange shaded region $0.01<\epsilon<0.17$) depending on the initial line mass in comparison to the edge formation (green solid line). Both timescales depend on the boundary density $\rho_b= 1.92 \times 10^{-20}$\gccm. But the plot is self-similar for different $\rho_b$. For $f<0.7$ the edge effect is faster than perturbations grow inside the filament, for more massive filaments perturbations could form slightly before the edges start to collapse.} 
	\label{fig:perturbations}
  \end{figure}
  
  To compare the formation of the edges to the general core formation inside a filament due to perturbations the two timescales are depicted in Figure \ref{fig:perturbations}. Following the analysis done in \citet{Heigl2020} the time dependent criticality of an infinitely long filament with perturbations of strength $\epsilon$ and growth timescale of the dominant mode $\tau_{\text{dom}}$ is given by:
  \begin{align}
    f_{\text{max}} (t) = f \left[1 + \epsilon \cdot \exp \left( \frac{t}{\tau _{\text{dom}}} \right) \right] \stackrel{!}{=} 1
  \end{align}
  Again, we consider a core to be formed for $f$ reaching a value of $1$ which leads to a core formation timescale due to perturbations of:
  \begin{align}
    t_{\text{pert}} = \tau _{\text{dom}} \log \left[ \left( \frac{1}{f} -1 \right) \frac{1}{\epsilon} \right]
  \end{align}
  The perturbation strength $\epsilon$ was set to $0.09$ according to observations by \citet{Roy2015} and $\tau _{dom}$ was calculated using the fourth-order polynomial function presented in \citet{Fischera2012} (Appendix E) based on \citet{Nagasawa1987}. Figure \ref{fig:perturbations} show the resulting collapse timescale due to perturbations depending on the initial criticality as orange dashed line ($\epsilon=0.09\pm0.08$, light orange shaded region) and in comparison the timescale of the edge effect (Equation \ref{eq:CollapseTime1.0}) as solid green line. Both timescales depend on the boundary density which was set to $\rho_b= 1.92 \times 10^{-20}$\gccm, the same as in our simulations. Although the absolute timescale varies when changing $\rho_b$ the trend stays the same: Below values of $f<0.7$ the formation of the edge effect is faster than the core formation timescale, above perturbations could grow faster.


\section{Parallel merger}
\label{subsec:ConditionsForAMerger}

  \subsection{Initial conditions}

\begin{figure}
	\centering
	\includegraphics[width=1.0\columnwidth]{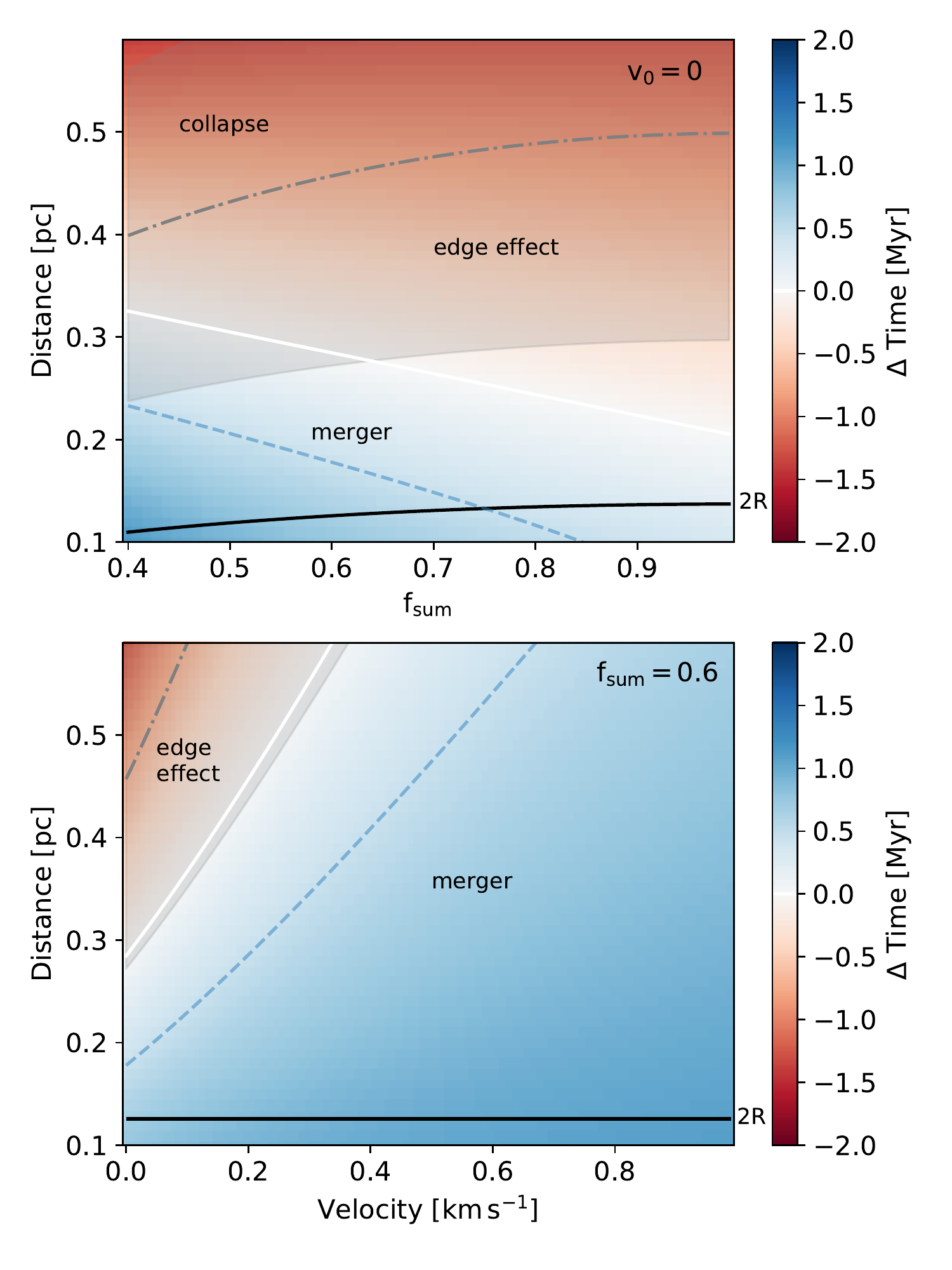}
	\caption{Time difference $\Delta t$ (Equation \ref{eq:DeltaT}) between merging and edge effect timescale for different initial conditions. The top panel shows the summed line-mass and distance of the two filaments (with $v_0=0$). The lower panel shows the distance and velocity for $f_{\text{sum}} = 0.6$. The white line indicates the conditions where the merging time equals the core-formation time at the edges. The dashed blue line indicates the initial distance for which filaments have enough time to form over densities of $f=0.5$ before merging. Thus, if the over-densities overlap supercritical regions are immediately created in the merged filament. The black line indicates two times the radius of the initial filament. Below this line filaments already overlap in the beginning. The region between the white and the black line (and more distinct the blue dashed line) give initial conditions where the filament merger is possible without forming supercritical regions. An initial velocity makes a mergers without supercritical densities more probable. The dashed dotted grey line gives the $t_{\text{col}}=t_{\text{merger}}$, Equation \ref{eq: Clarke} \citep{Clarke2015}, with an aspect ratio of $A=8$ (the grey shaded area gives the condition for $4<A<12$), above this we would expect a filament to collapse faster than it can merge.}
	\label{fig:InitialVelocities}
  \end{figure}

  With the analytic formulation of the merging and edge effect timescales it is now possible to calculate the conditions for the merging of two filaments without a dominant `edge effect'. In order to do so, we calculate the time difference between the merger (Equation \ref{eq: merging_time}) and the collapse (Equation \ref{eq: emp. model}) as function of the sum $f_{\text{sum}}=f_1+f_2$ the initial distance and the initial velocity:
  \begin{align}\label{eq:DeltaT}
    \Delta t = t_{\text{edge},1.0}-t_{\text{merger}}
  \end{align}
  The result is given in Figure \ref{fig:InitialVelocities}. 
  In the top panel $\Delta t$ (colour scale) is given as function of distance and $f_{\text{sum}}$. Here, only $f_1=f_2$ was considered. Otherwise, only the edge effect timescale of the heavier filament plays a role. In the lower panel it is given for $f_1=f_2=0.3$ depending on distance and the initial velocity. The negative (red) values belong to the regime where the filament forms cores faster than it can merge. Below the white line, which indicates the point where the timescales are equal, a merger is possible before the filaments from cores (blue regime). The dashed blue line gives the timescale needed for the filaments to form overdensities with values of $f=0.5$. Thus, the merged filament is immediately supercritical if these regions overlap. The merger of two filaments is limited by the grey dashed dotted line which is given by $t_{\text{col}}=t_{\text{merger}}$ (Equation \ref{eq: Clarke}, with a representative aspect ratio of $A=8$) whereas the grey shaded region gives the times for $4<A<12$. Above this line the filament collapses faster than it can merge. Thus, no merger is possible in that regime. The black line indicates the size of two times the radius of the initial filament. Below these values the filaments are already overlapping initially. Therefore we are only interested in the area above. In summary, for a symmetric merger of initially resting filaments, filaments initially have to lie between the black and the dashed blue lines (for overlapping end regions) or solid white lines (for not overlapping end regions) in order to merge before core collapse which appears rather unlikely. In the lower panel $\Delta t$ is depicted as function of velocity and distance for filaments of $f_1=f_2=0.3$. In this case a much wider range of distances is allowed for filament mergers, as the window widens up with increasing velocity. This is also the case for other summed line-masses, as depicted in Appendix \ref{ap:velocity} for $f_{\text{sum}}=0.4$ and $f_{\text{sum}}=0.8$. In general, the influence of the initial velocity is much stronger for lighter than for more massive filaments. Nevertheless, having an initial velocity makes a merger much more probable.

  This shows that symmetric filaments cannot easily merge, but need to fulfill strong initial conditions before the edge effect takes over: The distances between the filaments have to be small ($<0.4$\pc) and their relative velocities have to be high ($>0.3
  $\kms), conditions found among small scale filaments.

  \subsection{Simulation of parallel merges}

  \begin{figure}
	\centering
	\includegraphics[width=1.0\columnwidth]{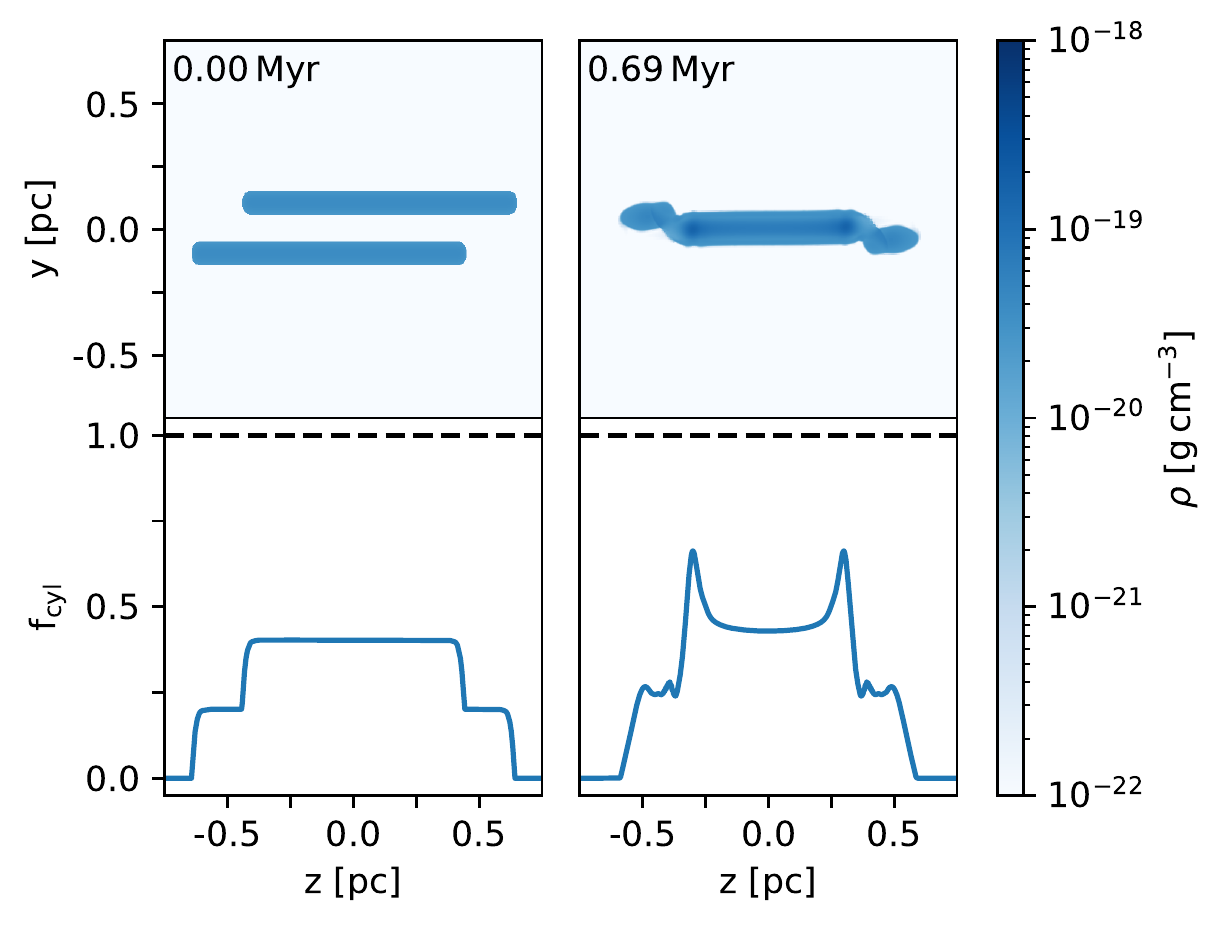}
	\caption{Same arrangement as in Figure \ref{fig:Simulation_Snapshots_collapse}. Two filaments with a critciallity of $f=0.2$, $d_0=0.2$\pc\ and $v_0=0.2$\kms\ are simulated with a translation along the z-axis. As expected they merge without creating supercritical regions.} 
	\label{fig:versetzt}
  \end{figure}

  By choosing the right initial conditions a merger without overcritcial cores can be simulated. An example is given in Figure \ref{fig:versetzt}. The initial conditions are shown in the left panel as density slice (top) and criticality (bottom). The filaments have a critciality of $f=0.2$ each, a relative velocity of $v_0 = 0.4$\kms\ and an initial distance of $d_0=0.2$\pc. Corresponding to Figure \ref{fig:App} (left hand side) the filaments should merge without forming supercritical regions which is what happens (see Figure \ref{fig:versetzt} bottom right). Because of the shift, two density peaks seem to evolve at the points where the edge overlaps with the other filament which could be a mechanism to form cores away from the edge of the new filament. Additionally, due to the shift, angular momentum is injected into the system which could generate accretion discs. We will explore these processes in details in a subsequent paper.

\section{Discussion} \label{sec:discussion}

  In this study we focused on the merging behavior of parallel aligned filaments. This is of course a special case which seems rather unlikely at first glance. Although the merging time was calculated for the parallel case, it should also give a good first approximation for inclined filaments.

  We simulated isothermal filaments without feedback, turbulence and winds. This isolated scenario could be linked to a young and quiet surrounding. Nevertheless, investigations by \citet{Hacar2018} already showed that the large scale feedback has no significant influence on the statistical properties of the fibers/filaments in Orion when comparing regions with low and high feedback. Thus, we do not expect large deviations for our model due to feedback. 

  We saw that the probability for filaments to merge is much higher for filaments with relative initial velocities. \citet{Hacar2013} already detected relative velocities up to $2.4$\kms in filaments of L1495/B213. However, if the initial velocity of the filaments is not oriented along their connecting line but with an inclination $\alpha$, too high velocities could lead to the escape of the filament out of the gravitational potential. The velocity necessary to escape the gravitational attraction is given by the escape velocity. For a configuration similar to the ones we simulated the escape velocity is of the order of $\sim 1$\kms. Filaments with higher velocities will only merge if their trajectories cross which is only the case for small angles. This results in a rather low probability for filaments to merge with relative velocities higher than the escape velocity. Further investigations have to show what else constrains relative velocities of merging filaments. Furthermore, high relative velocities could also destroy the filaments in the process of the merger. 

  We limited the effects of accretion by simulating a low density environment. Accretion could affect both the merging and collapse timescale as it increases the filament line-mass, exerts a ram pressure and induces turbulence in the filaments \citep{Clarke2015, Heigl2018_a, Heigl2020}. 

  The filaments were set-up as idealized cylinders with a constant line-mass. However, density gradients inside the filaments can influence the local merging timescale, according to Equation \ref{eq: merging_time} the merging time is shorter for larger $f$. This effect can lead to even more complex structures. Imagine a density minimum in the center of the filaments (along the z-axis): the ends of the filaments would merge first and ring like structures could from. Moreover, also the timescale and behavior of the edge effect would change since density enhancements are already present in the filament.

\section{Conclusions} \label{sec:conclusion}

  We provide an analytic model to describe both, the merger of filaments and the first phase of filament collapse due to the `edge effect'. Simulations show that the used approximations are reasonable and fit the predictions well. To conclude, the main points are as follows: 

  As the timescale of merging and collapsing filaments are of the same order of magnitude, the initial conditions determine how the resulting structure would look like. Three different outcomes are possible: The resulting filament has no supercritical cores at the edges, the edge effect dominates the resulting structure or the filaments collapse entirely before they can merge. The outcome mostly depends on the initial velocity, the initial line-mass and the initial distance of the filaments. For high velocities ($>0.3$\kms) and small distances ($<0.4$\pc) the probability for a symmetric filament merger is highest.

\section*{Acknowledgements}

  This research was supported by the Excellence Cluster ORIGINS which is funded by the Deutsche Forschungsgemeinschaft (DFG, German Research Foundation) under Germany's Excellence Strategy – EXC-2094 – 390783311. We thank the CAST group and especially Alvaro Hacar for helpful discussions and comments.

\section*{Data Availability}
 

The data included in this study will be made available on request.



\bibliographystyle{mnras}
\bibliography{literature_initialConditions} 




\newpage
\onecolumn

\appendix

\section{Solving the velocity integral} \label{ap: Calculation of the merging timescale}

  To calculate the merging time the ODE $\frac{\mathrm{d}d}{\mathrm{d}t}=v$ (velocity is given by Equation \ref{eq:velocity}) has to be solved, with initial condition $t_0=0$
  \begin{align}
    t &= - \int_{d_0}^{d} \frac{\mathrm{d}d'}{\sqrt{4G(\mu_1+\mu_2)\log\left(\frac{d_0}{d'}\right)+v_0^2}}
  \end{align}
  Substitution $u=\frac{d_0}{d'}$, $\frac{\mathrm{d}u}{\mathrm{d}d'}=-\frac{d_0}{d'^2}=-\frac{u^2}{d_0}$
  \begin{align}
    t = \int_{1}^{d_0/d}\mathrm{d}u \; \frac{d_0}{u^2} \left( 4G(\mu_1+\mu_2)\log(u)+v_o^2 \right)^{-\frac{1}{2}}
  \end{align}
  Substitution $x=\sqrt{4G(\mu_1+\mu_2)\log(u)+v_0^2}$
  \begin{align}
    &\frac{\mathrm{d}x}{\mathrm{d}u} = \frac{2G(\mu_1+\mu_2)}{u\sqrt{4G(\mu_1+\mu_2)\log(u)+v_0^2}}\\
    &u = \exp \left( \frac{x^2-v_0^2}{4G(\mu_1+\mu_2)} \right)
  \end{align}
  Inserting this, with $v_1 = \sqrt{4G(\mu_1+\mu_2)\log(\frac{d_0}{d})+v_0^2}$:
  \begin{align}
    t= \frac{d_0}{2G(\mu_1+\mu_2)} \int_{v_0}^{v_1}\mathrm{d}x \; \exp \left( -\frac{x^2-v_0^2}{4G(\mu_1+\mu_2)} \right)
  \end{align}
  Substitution $h = \frac{x}{\sqrt{4G(\mu_1+\mu_2)}}$, $\frac{\mathrm{d}h}{\mathrm{d}x}=\frac{1}{\sqrt{4G(\mu_1+\mu_2)}}$
  \begin{align}
    t =&\ \frac{d_0}{\sqrt{G(\mu_1+\mu_2)}} \exp \left( \frac{v_0^2}{4G(\mu_1+\mu_2)} \right) \cdot \int_{v_0/\sqrt{4G(\mu_1+\mu_2)}}^{v_1/\sqrt{4G(\mu_1+\mu_2)}}\mathrm{d}h \; \exp \left( -h^2 \right) \\
    =&\ \sqrt{\frac{\pi}{4G(\mu_1+\mu_2)}} d_0
    \exp \left( \frac{v_0^2}{4G(\mu_1+\mu_2)} \right) \left[ \mathrm{erf} \left( \frac{\sqrt{4G(\mu_1+\mu_2)\log \left( \frac{d_0}{d} \right) +v_0^2}}{\sqrt{4G(\mu_1+\mu_2)}} \right) - \mathrm{erf} \left( \frac{v_0}{\sqrt{4G(\mu_1+\mu_2)}} \right) \right] 
  \end{align}
  Finally, the time it takes to get from $d_0$ to a certain $d$ is given by (which is the resulting Equation \ref{eq: t merger t(r)}):
  \begin{align}
    t=&\ \sqrt{\frac{\pi}{G(\mu_1+\mu_2)}} \cdot \frac{d_0}{2} \cdot \exp \left( \frac{v_0^2}{4G(\mu_1+\mu_2)} \right) \cdot \left[ \mathrm{erf} \left( \sqrt{\log \left( \frac{d_0}{d} \right) + \frac{v_0^2}{4G(\mu_1+\mu_2)} } \right) -\mathrm{erf} \left( \frac{v_0}{\sqrt{4G(\mu_1+\mu_2)}} \right) \right]
  \end{align}

\section{Initial conditions for filament mergers depending on velocity} \label{ap:velocity}

  \begin{figure}
	\centering 
	\includegraphics[width=0.49\textwidth]{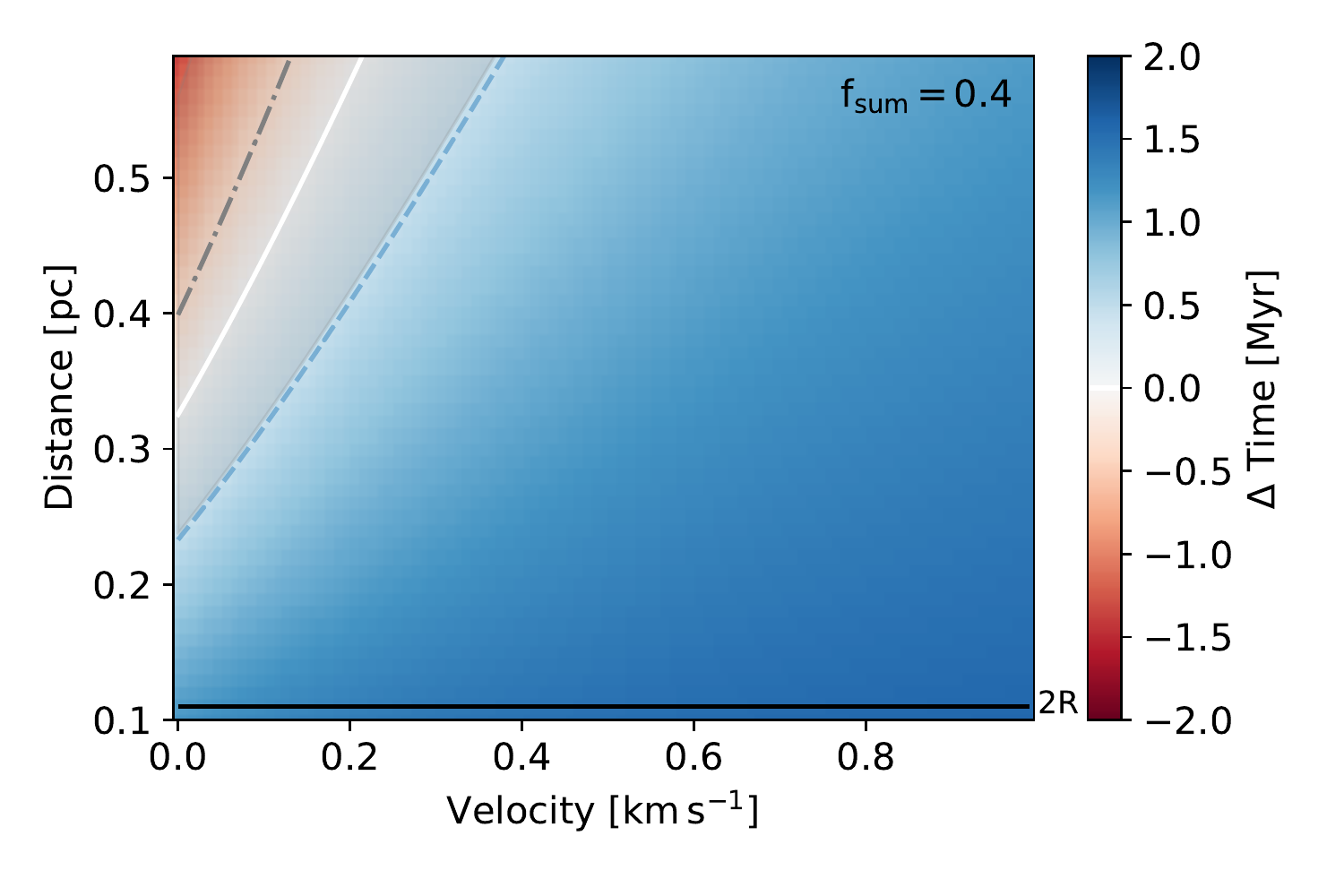}
	\includegraphics[width=0.49\textwidth]{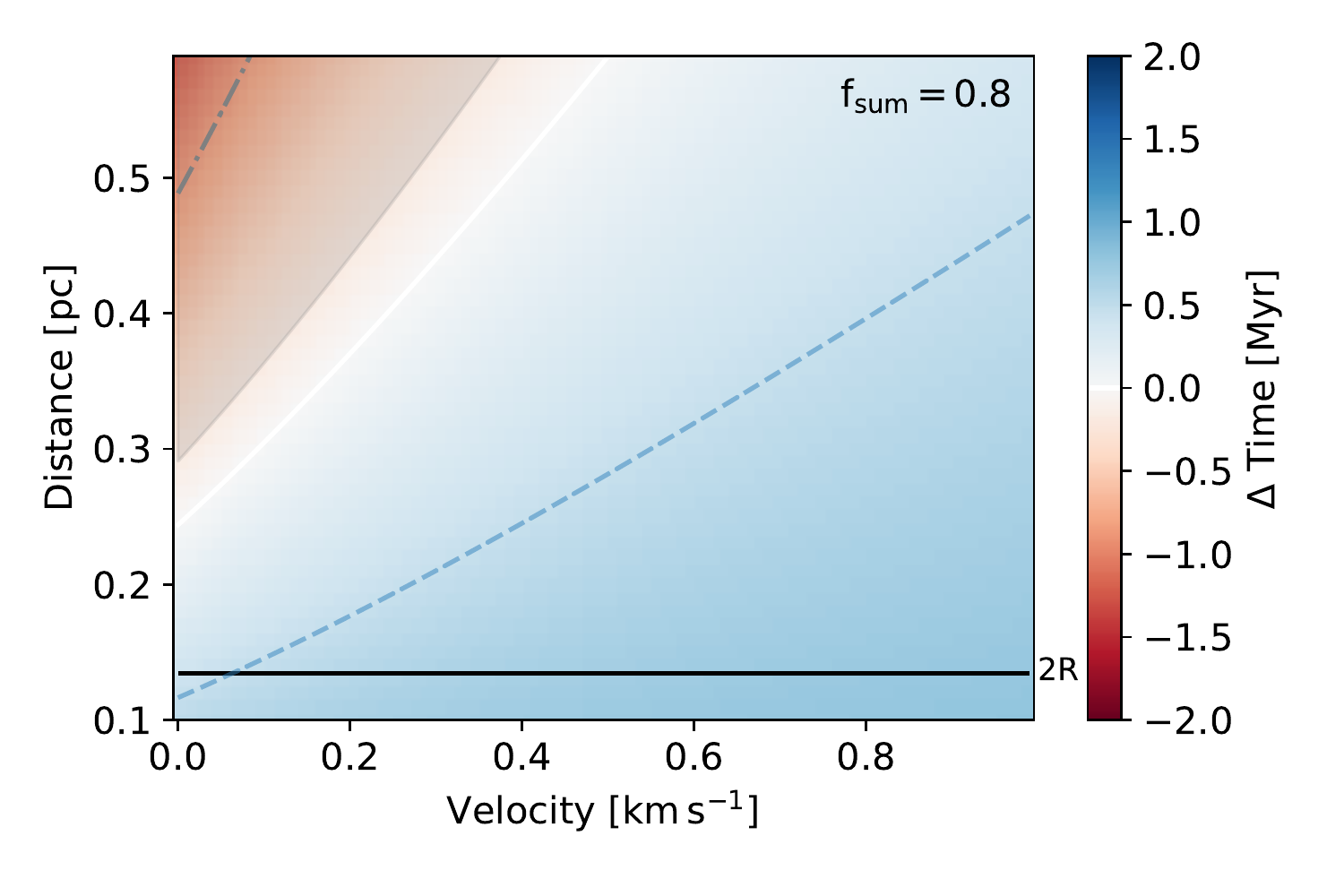}
	\caption{$\Delta t$ as defined in Equation \ref{eq:DeltaT} depending on diastance and velocity for different cirtialities $f_{\text{sum}}=0.4$ and $f_{\text{sum}}=0.8$ (a more detailed description is given in Figure \ref{fig:InitialVelocities}). This shows that in both cases an initial velocity makes a merger more probable. Though, the influence on the lower line mass filament is stronger.} 
	\label{fig:App}
  \end{figure}


\bsp	
\label{lastpage}
\end{document}